\documentclass[11pt,twoside]{article}

\usepackage[utf8]{inputenc}
\usepackage{epsfig,amsfonts,color}
\usepackage{mathrsfs} 
\usepackage{amsmath, bbm, mathabx, amsthm}
\usepackage[font=small]{caption}
\usepackage[normalem]{ulem}
\usepackage{amssymb, palatino, geometry,url}
\usepackage[colorlinks=true,linkcolor=blue,citecolor=blue,urlcolor=blue]{hyperref}
\usepackage{subcaption}
\usepackage{multirow}

\usepackage{graphicx} 

\usepackage{fancyhdr}
\usepackage{lineno}
\usepackage{float}
\usepackage[section]{placeins}
\usepackage[table]{xcolor}

\title{Topological Data Analysis for High-Dimensional Dynamic Process Monitoring}

\author{Angan Mukherjee$^1$, Tyler A. Soderstrom$^2$, Michael J. Kurtz$^2$, and Victor M. Zavala$^{1}$\thanks{Corresponding Author: zavalatejeda@wisc.edu}\\[0.6em]
    {\small $^1$Department of Chemical \& Biological Engineering,}\\
    {\small \;University of Wisconsin-Madison, 1415 Engineering Drive, Madison, WI 53706, USA}\\
    {\small $^2$ExxonMobil Technology and Engineering, 22777 Springwoods Village Pkwy, Spring, TX 77389, USA}
    }
\date{}

\begin{document}

\maketitle

\begin{abstract}
Real-time process monitoring requires methods that extract actionable information from high-dimensional time-series data. In this work, we present a new approach for process monitoring that combines tools of topological data analysis (TDA) and machine learning. In the proposed approach, we represent multivariate time-series data as manifolds and use topological descriptors to summarize the structure of such data; we then use a neural ordinary differential equation to learn the dynamic evolution of the topological structure of the system. Using real data from an industrial process, we show that this trajectory-based event detection approach is effective at detecting diverse types of events. We contrast this approach against reconstruction-based approaches such as principal component analysis and autoencoders and against a trajectory-based approach that uses Koopman autoencoders. 
\end{abstract}

\section{Introduction} \label{sec:introduction}

Chemical processes generate complex multivariate time-series data that comprise sensor measurements for hundreds, if not thousands, of variables. These datasets encode rich information about the dynamic state of the process and are central to process monitoring, fault detection, safety analysis, and decision-making. However, detecting normal/abnormal events from these high-dimensional datasets is challenging; this is often due to the highly coupled, nonlinear, and multi-timescale nature of chemical processes and due to the presence of high measurement noise. As a result, there continues to be a need for {\textit {process monitoring frameworks}} that are scalable, interpretable, and can detect different types of events (e.g., major and minor) \cite{park2020review,nor2020reviewfault}.

The development of robust fault detection and diagnosis tools has received significant attention across diverse industrial applications. Broadly, these methods may be categorized as {\textit{reconstruction-based}} or {\textit{trajectory-based}} approaches, depending on whether event signatures are inferred from static reconstruction mismatch or from temporal evolution of system features. 

Several statistical and data-driven event-detection methods have been developed and used for process monitoring and fault detection; these include: principal component analysis \cite{venkat2003review1} (PCA), convolutional neural networks \cite{jiang2023tdacnn,gonzalez2025pastillation} (CNN), support vector machines \cite{venkat2003review3,chiang2004fault}, Fisher discriminant analysis \cite{chiang2004fault}, and autoencoders \cite{qian2022aefault} (AE). Among these, PCA and AE are representative {\textit{reconstruction-based}} approaches that compress high-dimensional data into a lower-dimensional latent space and detect faults/events via reconstruction errors relative to nominal operation. Although such approaches are often effective at identifying major or persistent disturbances, they can struggle to detect subtle, minor, or dynamically evolving instabilities embedded in noisy multivariate process datasets \cite{nor2020reviewfault}. {\textit{Trajectory-based}} monitoring approaches explicitly exploit the temporal structure of multivariate process data. Koopman autoencoders (KAE), for instance, learn nonlinear transformations under which the underlying latent-space dynamics evolve {\textit{linearly}} \cite{wang2025koopAE}. The resulting latent-space evolution provides a useful benchmark for comparing dynamic data representations for multivariate process monitoring and event detection \cite{park2026koopAE}.

\begin{figure}[htbp]
    \centering
    \includegraphics[width=0.99\textwidth]{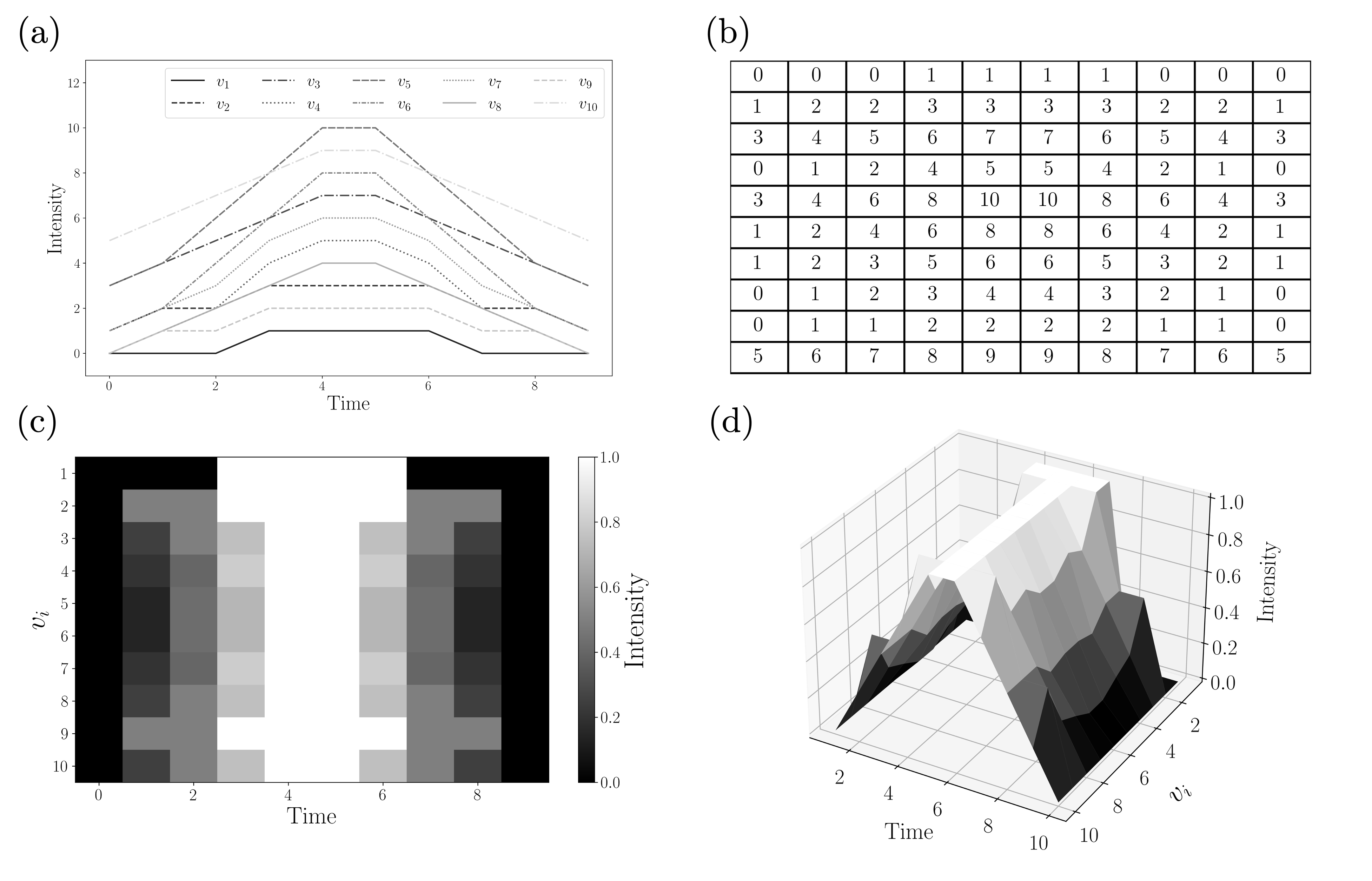}
    \caption{Different representations of multivariate time-series data. (a) Raw time-series trajectories for process variables $v_i$. (b) A matrix representation of the same data, where rows correspond to variable indices and columns to time steps. (c) A heatmap visualization of the matrix, forming a 2D manifold whose pixel intensity reflects normalized process variables. (d) A 3D surface / field obtained by projecting intensities onto a third (vertical) axis, highlighting geometric structure (shape) in the data.}
    \label{fig:data_rep}
\end{figure}

A fundamental issue underlying event detection approaches is {\textit{data representation}}. For example, multivariate time-series data are typically represented as vectors and matrices, as needed by tools from statistics, linear algebra, and machine learning. However, modern tools of machine learning and data analysis can also process datasets that are represented as images, manifolds, point clouds, graphs, or networks \cite{smith2021tda,jiang2021convolutional,jiang2024convolutional}. The data representation used influences the technique used for extracting information and the type of information extracted. Figure \ref{fig:data_rep} illustrates this idea for multivariate time-series data, showing how trajectories may be reorganized into matrix, image, and manifold representations.

Topological data analysis (TDA), particularly persistent homology \cite{smith2021tda}, provides a natural framework for exploiting such structured data representations. In recent years, TDA has found success in applications spanning neuroscience, materials science, and network analysis \cite{caputi2021tdaneuro,guo2017tdamanufac,zheng2023percept}. By representing  data as graphs, networks, simplicial complexes, or manifolds, TDA enables dimensionality reduction into {\textit{informative}} lower-dimensional spaces by computing topological descriptors (e.g., persistence diagrams, Betti numbers, and the Euler characteristic (EC)) that aim to quantify the shape/structure of data \cite{smith2021ecpaper,smith2021tda,laky2024fastec,zavala2023MLpers}. TDA also offers useful mathematical properties, such as robustness to noise, and invariance under certain classes of perturbations such as scaling, rotation, and deformations \cite{smith2023ecmd}. Moreover, TDA computations are more scalable than machine learning techniques, such as convolutional neural networks; this is because TDA does not require training procedures \cite{laky2024fastec}. These features make TDA a promising approach to process monitoring \cite{ravishanker2019tdatime,zheng2024tdamultivar,yaagoubi2023tdamultivar}.

A {\textit{key}} indicator of the onset of faults or events in multivariate time-series data is change in dynamic directionality (i.e., changes in the underlying dynamical structure of the system). In the absence of high-fidelity mechanistic models, ML models offer a powerful approach to capture such complex dynamical signatures directly from data \cite{mukherjee2025pcml}. In particular, dynamic ML models such as recurrent neural networks \cite{mukherjee2023hybserpar} (RNNs) and neural ordinary differential equations \cite{thompson2025pcnode} (NODEs) can be used to learn the temporal evolution of process from data. This motivates the idea that, instead of learning the evolution of the original high-dimensional system directly, one may learn the evolution of a lower-dimensional latent representation that captures the evolution of the process. This approach has been used for learning the dynamics of a system in a latent space using Koopman operators and autoencoders \cite{wang2025koopAE}. TDA also offers a compact latent representation of the system that captures its topological structure; as such, one can envision developing ML dynamical models of the topological structure of the system. The key difference with autoencoder approaches is the interpretation of the latent space and the type of information contained in such space.

In this work, we present a computational framework for multivariate time-series process monitoring that combines TDA and NODE. We benchmark the proposed against common reconstruction-based and trajectory-based event-detection approaches (PCA, AE, and KAE). The proposed framework represents the multivariate time-series data as {\textit {manifolds}}, extracts topological descriptors through filtration operations, and uses a NODE model to learn how these descriptors evolve over time. Through this approach, we show that {\textit {significant deviations}} in the time derivatives of topological descriptors correspond to the onset of distinct events or unstable operating regions in the process. By comparing against other monitoring approaches on real industrial datasets, we demonstrate that the proposed framework provides interpretable, robust, and scalable signatures for detecting both major events and subtle minor events.

The rest of the paper is organized as follows. Section \ref{sec:method} introduces reconstruction-based and trajectory-based monitoring approaches considered in this work and presents the proposed TDA-NODE framework. Section \ref{sec:case_study} describes the industrial dataset and the representative events considered for benchmarking. Section \ref{sec:results} compares the performance of the different monitoring approaches. Section \ref{sec:conclusions} summarizes the main findings and discusses directions for future work.

\section{Approaches to Process Monitoring} \label{sec:method}

We introduce approaches for event detection in multivariate time-series data. We organize the methodology around a couple of broad classes of approaches: {\textit{reconstruction-based}} methods, which detect events through deviations in reconstruction errors relative to nominal operation, and {\textit{trajectory-based}} methods, which detect events through changes in the temporal evolution of a latent space. We consider principal component analysis (PCA) and autoencoders (AE) as reconstruction-based approaches, and Koopman autoencoders (KAE) together with the proposed topological data analysis (TDA) framework as trajectory-based approaches. For the TDA approach, we discuss the {\textit{representation}} of multivariate time-series data as manifolds, {\textit{extraction}} of information in the form of topological descriptors, and {\textit{modeling}} of temporal evolution using differential machine-learning models. 
\\

We begin by considering a multivariate time-series collected from a process with $n \in \mathbb{Z}_+$ measured variables over a monitoring horizon $[0,T]$, where $T \in \mathbb{R}_+$. Let
\begin{equation}
{\bf x}(t) \in \mathbb{R}^{n}, \qquad t \in [0,T]
\end{equation}
denote the process measurement vector at time $t$. In practice, measurements are collected at discrete time instants:
\begin{equation}
0 \le t_1 < t_2 < \cdots < t_M \le T
\end{equation}
where $M \in \mathbb{Z}_+$ is the total number of time samples. The sampled dataset is represented by the matrix
\begin{equation}\label{data_mat}
{\bf X}
=
\begin{bmatrix}
{\bf x}(t_1)^{\top}\\
{\bf x}(t_2)^{\top}\\
\vdots\\
{\bf x}(t_M)^{\top}
\end{bmatrix}
\in \mathbb{R}^{M \times n}
\end{equation}

To detect events, we process data using a moving window approach. Let $K \in \mathbb{Z}_+$ denote the number of windows. The $k$-time window is given by:
\begin{equation}
W_k = [\tau_k^{s},\tau_k^{e}] \subseteq [0,T] \qquad k=1,\dots,K
\end{equation}
where $\tau_k^{s},\tau_k^{e} \in [0,T]$ are the start and end times of the window and thus $\tau_k^{s}<\tau_k^{e}$. Additionally, the windows are chosen so that consecutive windows overlap. For each window $W_k$, we define the index set:
\begin{equation}
\mathcal{I}_k = \{ m \in \{1,\dots,M\}: t_m \in W_k \}
\end{equation}
and let $M_k = |\mathcal{I}_k|$ denote the number of samples contained in the window. The sampled data restricted to window $W_k$ is then represented by the matrix
\begin{equation}
{\bf X}_k
=
\begin{bmatrix}
{\bf x}(t_{m_1})^{\top}\\
{\bf x}(t_{m_2})^{\top}\\
\vdots\\
{\bf x}(t_{m_{M_k}})^{\top}
\end{bmatrix}
\in \mathbb{R}^{M_k \times n}
\qquad \{m_1,\dots,m_{M_k}\} = \mathcal{I}_k
\end{equation}

We associate a representative time $\tau_k \in W_k$ within each window by using the start time of the window: 
\begin{equation}
\tau_k = \tau_k^{s}, \qquad k=1,\dots,K
\end{equation}
The moving-window construction is used to represent the full monitoring horizon as a sequence of $K$ overlapping data windows. Thus, the process monitoring approaches considered in this work are not applied to a single isolated window, but rather across the entire sequence $\{{\bf X}_k\}_{k=1}^{K}$ spanning the full time horizon. 

In reconstruction-based methods, reconstruction errors are evaluated on a per-window basis, yielding a sequence of error metrics over the monitoring horizon. In trajectory-based methods, each window is mapped to a latent variable ${\bf z}_k \in \mathbb{R}^{r}$, where $r \in \mathbb{Z}_{+}$ is the dimension of the latent space, and event signatures are inferred from the temporal evolution of the sequence $\{{\bf z}_k\}_{k=1}^{K}$. This moving-window implementation provides a common mathematical basis for comparing reconstruction-based and trajectory-based approaches. All codes required to implement the reconstruction- and trajectory-based process monitoring approaches discussed in this paper, along with a representative test-case dataset, are available at: \url{https://github.com/zavalab/ML/tree/master/TDA_Process_Monitoring}.

\subsection{Principal Component Analysis (PCA)} \label{sec:PCA_math}

PCA is a widely used reconstruction-based approach that projects each windowed data matrix ${\bf X}_k \in \mathbb{R}^{M_k \times n}$ into a lower-dimensional latent space of dimension $r \le n$. Let ${\bf P} \in \mathbb{R}^{n \times r}$ denote the loading matrix whose columns form an orthonormal basis for the principal subspace. The latent scores for window $k$ are given by
\begin{equation}
{\bf T}_k = {\bf X}_k {\bf P} \in \mathbb{R}^{M_k \times r}
\end{equation}
and the corresponding reconstructed data matrix is defined as:
\begin{equation}
\widehat{\bf X}_k = {\bf T}_k {\bf P}^{\top} = {\bf X}_k {\bf P}{\bf P}^{\top} \in \mathbb{R}^{M_k \times n}
\end{equation}
The reconstruction error for window $k$ is thus represented as the Q-statistic:
\begin{equation}
\varepsilon_k = \|{\bf X}_k - \widehat{\bf X}_k\|
\end{equation}
Large values of $\varepsilon_k$ indicate that the data in window $k$ departs from the nominal PCA subspace and are therefore used as indicators of events \cite{lieftucht2006pca}. The implementation of this approach for $k=1,\dots,K$ yields the sequence of reconstruction errors $\{\varepsilon_k\}_{k=1}^{K}$ over the full monitoring horizon.

\subsection{Autoencoders (AE)} \label{sec:AE_math}

AE approaches generalize PCA by learning nonlinear mappings between each windowed data matrix ${\bf X}_k \in \mathbb{R}^{M_k \times n}$ and a latent space of dimension $r$. Let
\begin{equation}
E_{\theta}: \mathbb{R}^{M_k \times n} \to \mathbb{R}^{r},
\qquad
D_{\phi}: \mathbb{R}^{r} \to \mathbb{R}^{M_k \times n}
\end{equation}
denote the encoder and decoder mappings, parameterized by trainable parameters $\theta$ and $\phi$, respectively. For each window $k$, the latent space variables and reconstructed data are given by:
\begin{subequations}\label{rep_AE}
\begin{align}
{\bf z}_k &= E_{\theta}({\bf X}_k) \in \mathbb{R}^{r}\\
\widehat{\bf X}_k &= D_{\phi}({\bf z}_k) \in \mathbb{R}^{M_k \times n}
\end{align}
\end{subequations}
The reconstruction error for window $k$ is hence defined as:
\begin{equation}
\varepsilon_k = \|{\bf X}_k - \widehat{\bf X}_k\|
\end{equation}
Similar to PCA, large values of $\varepsilon_k$ across the full time horizon indicate poor reconstruction of the incoming measurements by the learned nonlinear latent representation during monitoring and are therefore interpreted as signatures of departures from nominal operation.

\subsection{Koopman Autoencoders (KAE)} \label{sec:KAE_math}

KAE is a {\textit{trajectory-based}} extension of the conventional AE that imposes approximately linear latent dynamics across successive time windows. Using a encoder and decoder structure as in \eqref{rep_AE}, the latent space vector for window $k$ are defined by
\begin{equation}
{\bf z}_k = E_{\theta}({\bf X}_k) \in \mathbb{R}^{r}
\end{equation}
and the reconstruction is
\begin{equation}
\widehat{\bf X}_k = D_{\phi}({\bf z}_k) \in \mathbb{R}^{M_k \times n}
\end{equation}
The temporal evolution of the latent variables is modeled by
\begin{equation}
{\bf z}_{k+1} \approx {\bf A}{\bf z}_k,
\qquad
{\bf A} \in \mathbb{R}^{r \times r}
\end{equation}
where ${\bf A}$ is a learned linear operator. The corresponding latent dynamic increment is defined as:
\begin{equation}\label{error_KAE}
\Delta {\bf z}_k = {\bf z}_{k+1} - {\bf z}_k \approx ({\bf A}-{\bf I}){\bf z}_k
\end{equation}
where ${\bf I} \in \mathbb{R}^{r \times r}$ is the identity matrix. In KAE-based monitoring, event signatures are inferred from the magnitude of the temporal variation of the latent-space vector, quantified by the metric $\|\Delta {\bf z}_k\|$. The reasoning is that sharp increases in $\|\Delta {\bf z}_k\|$ indicate abrupt changes in the latent dynamics and are therefore interpreted as event signatures.

\subsection{TDA Framework} \label{sec:tda_descriptors}

The proposed TDA framework is a {\textit{trajectory-based}} monitoring approach that consists of three main components: (a) constructing a manifold representation from the multivariate time-series data, (b) extracting topological descriptors from this representation, and (c) learning the temporal evolution of the resulting descriptor sequence using a NODE model.

\subsubsection{Manifold Data Representation} \label{sec:data_rep}

The sampled multivariate time-series data are represented by the matrix ${\bf X} \in \mathbb{R}^{M \times n}$, as defined in \eqref{data_mat}, where $M$ is the number of sampled time instants and $n$ is the number of process variables, denoted by $v_i$, $i=1,\dots,n$. We define the transposed and normalized matrix
\begin{equation}
{\bf Y} = \mathcal{N}({\bf X}^{\top}) \in [0,1]^{n \times M}
\end{equation}
where $\mathcal{N}(\cdot)$ denotes a variable-wise normalization operator that scales each process variable independently to the interval $[0,1]$. 

The matrix ${\bf Y}$ defines a 2D manifold representation on the discrete grid:
\begin{equation}
\mathcal{G} = \{1,\dots,n\} \times \{1,\dots,M\}
\end{equation}
where the first coordinate indexes process variables and the second coordinate indexes sampled time instants. Thus, the intensity value at each grid point, i.e., ${Y}_{i,j} \in [0,1]$, denotes the normalized value $(\tilde{v})$ of process variable at the $j$-th sampled time instant.

\begin{figure}[!htbp]
    \centering
    \includegraphics[width=0.995\textwidth]{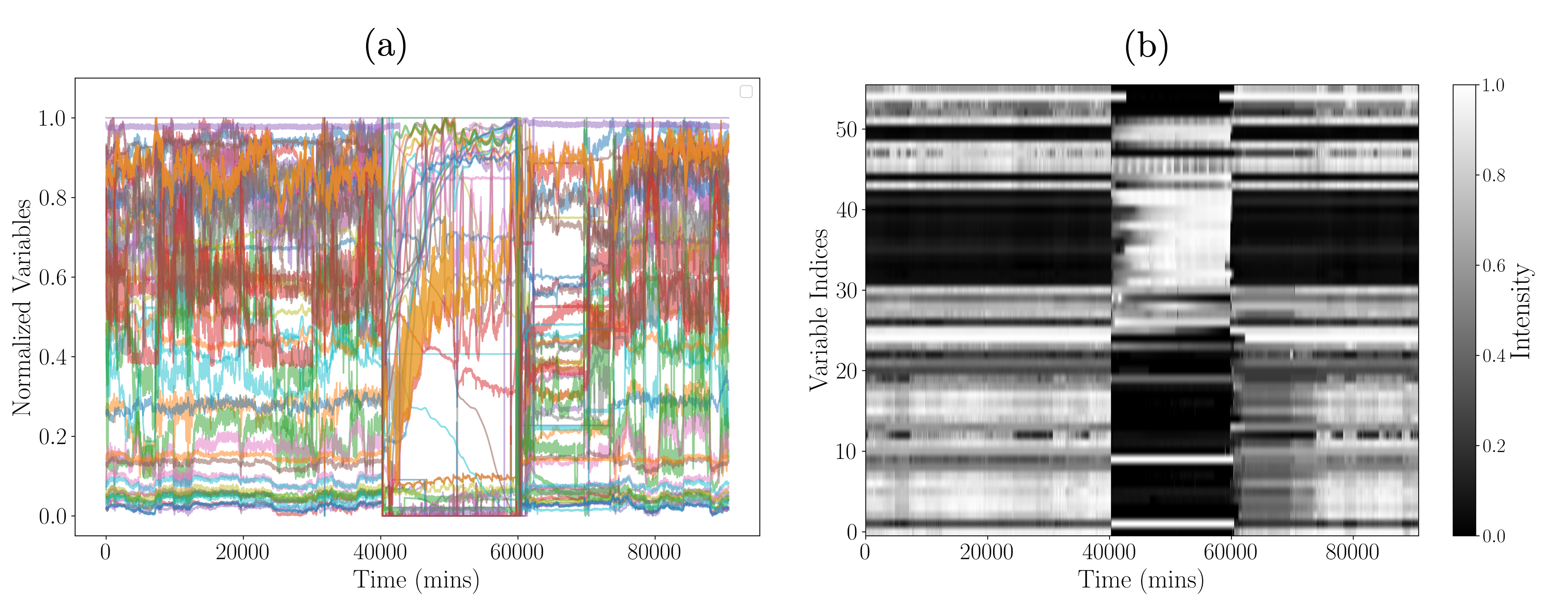}
    \caption{Different representations of a multivariate time-series dataset. (a) Raw normalized trajectories for $N=56$ variables over a monitoring horizon of approximately $90{,}000$ minutes. (b) 2D manifold representation of the same dataset, where rows correspond to variable indices, columns correspond to time indices, and grayscale intensity denotes the normalized value of each variable.}
    \label{fig:data_manifold}
\end{figure}

In the context of process monitoring, the manifold representation is restricted to each moving window $W_k$, yielding a sequence of window-specific manifolds associated with the matrices ${\bf X}_k$, $k=1,\dots,K$. Figure \ref{fig:data_manifold} illustrates this construction for a representative dataset with $N=56$ measured variables and a monitoring horizon of approximately $90{,}000$ minutes. Figure \ref{fig:data_manifold}a shows the raw normalized time-series trajectories of the process variables, whereas Figure \ref{fig:data_manifold}b shows the corresponding manifold representation. The manifold representation captures the high-dimensional trajectories in a form that makes spatiotemporal patterns more salient \cite{jeong2024manifold}, and hence is used as the basis for extraction of the topological descriptors introduced in the next subsection.

\subsubsection{Topological Descriptors} \label{sec:TDA_desc}

For each moving window $W_k$, the corresponding restriction of the manifold representation yields a window-specific matrix ${\bf Y}_k \in [0,1]^{n \times M_k}$ defined on the discrete grid
\begin{equation}
\mathcal{G}_k = \{1,\dots,n\} \times \{1,\dots,M_k\}
\end{equation}
We interpret ${\bf Y}_k$ as a scalar field $h_k : \mathcal{G}_k \to [0,1]$ defined by:
\begin{equation}
h_k(i,j) = ({\bf Y}_k)_{i,j},
\qquad
(i,j) \in \mathcal{G}_k
\end{equation}

Using this field, we construct a sublevel-set filtration indexed by a threshold $\ell \in [0,1]$. For each $\ell$, the corresponding sublevel set is defined as
\begin{equation}
\mathcal{S}_k(\ell) = \{(i,j)\in \mathcal{G}_k : h_k(i,j) \le \ell \}
\end{equation}
This sublevel set may be represented as a cubical complex induced by the active grid points at threshold $\ell$ \cite{smith2021ecpaper,huber2021superlevel}.

For each threshold $\ell$, we let
\begin{equation} 
\beta_p^{(k)}(\ell) \in \mathbb{Z}_{+} \cup \{0\},
\qquad p=0,1
\end{equation}
denote the $p$-th Betti number of the cubical complex associated with $\mathcal{S}_k(\ell)$. Here, $\beta_0^{(k)}(\ell)$ denotes the number of connected components and $\beta_1^{(k)}(\ell)$ denotes the number of one-dimensional holes. Because the manifold representation is 2D, the Euler characteristic corresponding to window $k$ and threshold $\ell$ is defined by
\begin{equation}\label{EC}
\chi_k(\ell) = \beta_0^{(k)}(\ell) - \beta_1^{(k)}(\ell),
\qquad \chi_k(\ell) \in \mathbb{Z}
\end{equation}
Additionally, the function $\chi_k : [0,1] \to \mathbb{Z}$ is referred to as the Euler characteristic (EC) curve associated with window $k$.

\begin{figure}[htbp]
    \centering
    \includegraphics[width=1\textwidth]{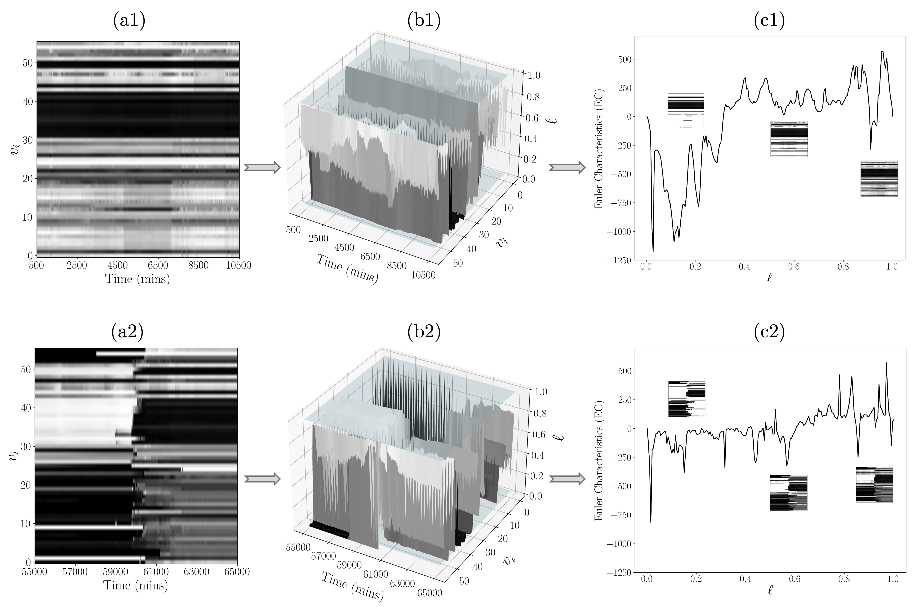}
    \caption{Extraction of topological descriptors from data matrices corresponding to two different time windows. For each window, (a1,a2) show the 2D manifold, (b1,b2) show the corresponding 3D field obtained by projecting intensities onto the vertical axis, and (c1,c2) show the EC curve obtained through sublevel-set filtration. The differences in the EC curves across windows indicate that the topology of the multivariate data evolves over time, motivating the analysis of the temporal directionality of EC descriptors for event detection.}
    \label{fig:filtration_EC}
\end{figure}

Figure \ref{fig:filtration_EC} illustrates this construction for two representative windows. Different windows typically yield different scalar fields and, consequently, different EC curves. This observation motivates our focus on the {\textit{directionality}} of the topological descriptors, i.e., on how the EC curves evolve across successive windows during process monitoring. The sequence $\{\chi_k\}_{k=1}^{K}$ therefore provides a time-resolved topological summary of the multivariate time-series data and forms the basis for dynamical modeling using a NODE model, which we dicuss next.

\subsubsection{Learning Topology Dynamics} \label{sec:node}

After extracting topological descriptors (EC curves) for each moving window, we seek to model how these descriptors evolve across the sequence of windows. This aims to capture how the structure of the dynamical system evolves over time. Several approaches in the literature focus on identifying deviations or transitions in {\textit{topological signatures}}, such as change-point detection \cite{zheng2023percept}, Wasserstein and bottleneck distances for comparing persistence diagrams \cite{clara2022wassersteintda}, and other TDA-based dissimilarity measures \cite{schieber2017dissimtda}. While such techniques are useful for quantifying differences between individual snapshots, they do not {\textit{explicitly}} model the temporal dynamics and directionality of the topological state of the process.

To this end, for the TDA framework, the latent variables are represented by topological descriptors extracted from each moving window, such as EC curves (and, more generally, Betti curves). Accordingly, let ${\bf z}_k \in \mathbb{R}^{r}$ denote the vector obtained by sampling the EC curve associated with window $k$ at $r \in \mathbb{Z}_{+}$ prescribed filtration thresholds. For notational simplicity, we use $\chi_k$ to represent the EC curve associated with the $k$-th window with representative time $\tau_k$, i.e., $\chi_k \equiv \chi(\tau_k)$. Specifically,
\begin{equation}
{\bf z}_k =
\begin{bmatrix}
\chi_k(\ell_1) & \chi_k(\ell_2) & \cdots & \chi_k(\ell_r)
\end{bmatrix}^{\top}
\in \mathbb{R}^{r}
\end{equation}
where $0 \le \ell_1 < \ell_2 < \cdots < \ell_r \le 1$. The latent topological state for window $k$ is identified with the finite-dimensional representation of the EC curve, i.e., ${\bf z}_k \equiv \chi(\tau_k)$.

To learn the temporal evolution of the sequence $\{{\bf z}_k\}_{k=1}^{K}$, we employ a continuous-time NODE model \cite{thompson2025pcnode}. Specifically, we assume that the latent topological state ${\bf z}(t) \in \mathbb{R}^{r}$ evolves according to
\begin{equation}\label{node}
\dot{\bf z}(t) = f_{\theta}({\bf z}(t)),
\qquad t \in [0,T]
\end{equation}
where $\dot{\bf z}(t) = \frac{d{\bf z}(t)}{dt}$, and $f_{\theta} : \mathbb{R}^{r} \to \mathbb{R}^{r}$ is a feedforward neural network (NN) parameterized by $\theta$. In this work, $f_{\theta}$ is implemented as a NN with one hidden layer of $64$ neurons and hyperbolic tangent activation function.

\begin{figure}[htbp]
    \centering
    \includegraphics[width=1\textwidth]{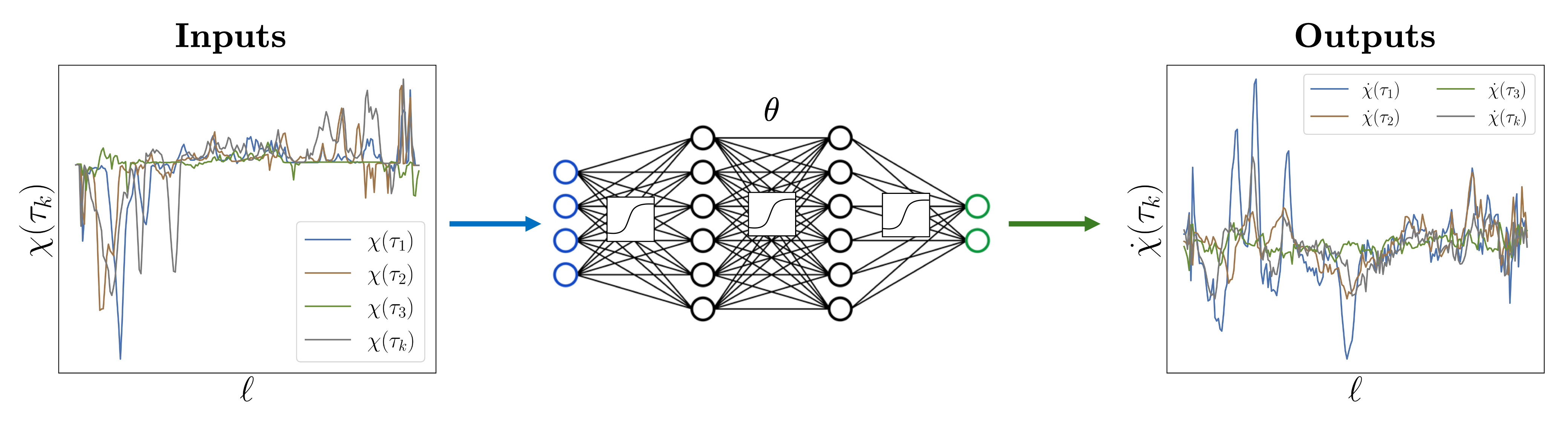}
    \caption{NODE architecture used to learn the temporal evolution of the topological descriptors. The input consists of EC curves $\chi(\tau_k)$ associated with successive moving windows, and the model predicts their corresponding temporal derivatives $\dot{\chi}(\tau_k)$. The learned directionality of the EC curves is subsequently used to identify events during process monitoring.}
    \label{fig:node}
\end{figure}

Given the sequence $\{(\tau_k,{\bf z}_k)\}_{k=1}^{K}$ associated with the moving windows, the NODE is trained to approximate the continuous-time evolution of the topological descriptor (here, EC curve) across successive windows. As illustrated in Figure \ref{fig:node}, the NODE takes $\chi(\tau_k)$ as inputs and predicts their instantaneous {\textit {rate of change}}. 

Such formulation enables the NODE to learn smooth trajectories in the topological space while detecting sharp deviations in their transients that represent the onset of events. Large values of the predicted derivative norm, $\|\dot{\chi}(\tau_k)\|$, indicate abrupt changes in the topological structure of the process and are therefore interpreted as event signatures during process monitoring.

\section{Case Study: Industrial Olefins Plant} \label{sec:case_study}

\begin{figure}[htbp]
    \centering
    \includegraphics[width=1\textwidth]{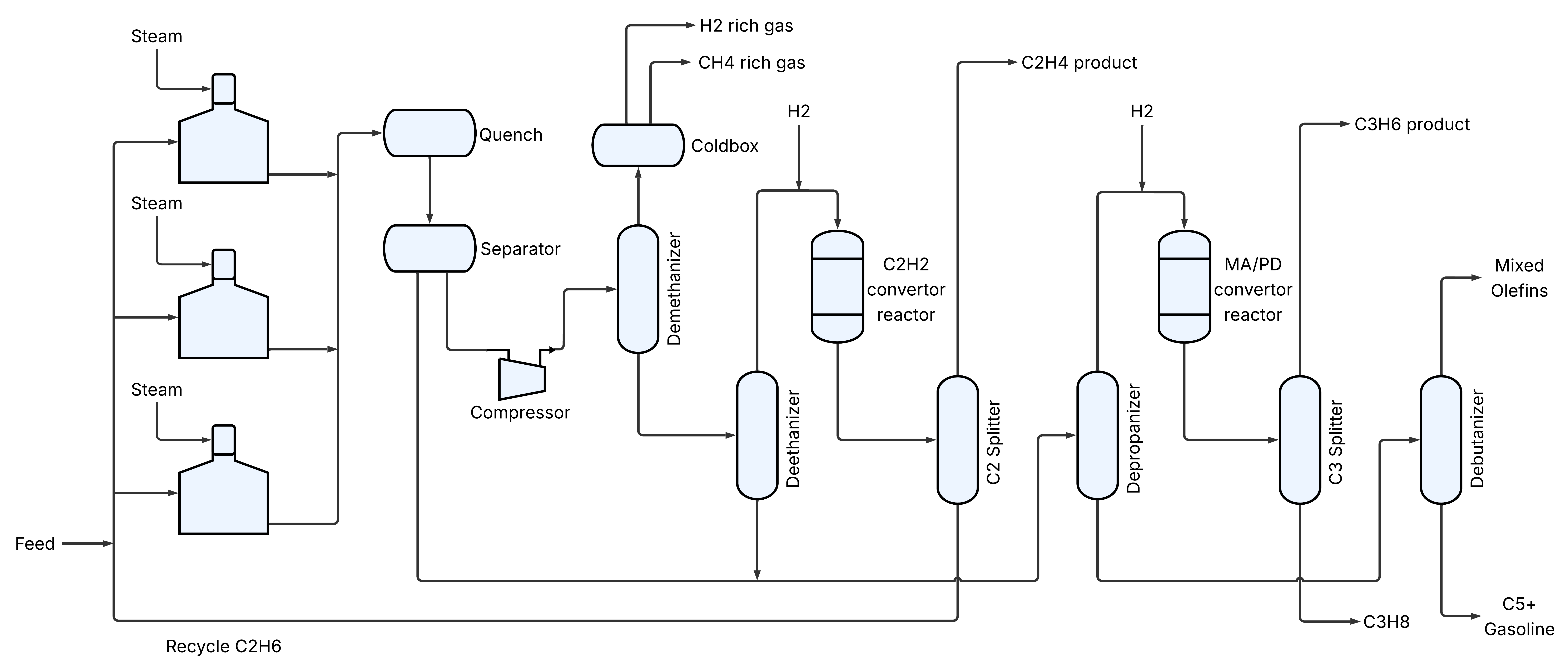}
    \caption{Schematic of industrial olefins plant. The plant converts hydrocarbon feedstocks into light olefins such as ethylene and propylene through steam cracking followed by downstream separation and purification. The process involves quenching, compression, and cryogenic distillation through units such as the demethanizer, deethanizer, and multiple splitter columns to separate hydrogen- and methane-rich streams.}
    \label{fig:case_study}
\end{figure}

To benchmark the monitoring approaches introduced in Sections \ref{sec:introduction} and \ref{sec:method}, we consider a real-world industrial olefins plant for which process data are provided. Figure \ref{fig:case_study} presents a schematic of a representative plant \cite{Siemens2018EthyleneGC}; this converts hydrocarbon feedstocks into light olefins, primarily ethylene and propylene, through thermal cracking followed by multistage separation and purification \cite{amghizar2017olefin}. Typically, steam cracking generates a complex mixture of hydrocarbons and byproducts that is rapidly cooled in a quench system to prevent undesired side reactions. The cooled stream is then routed through compression and a sequence of separation units, including a demethanizer, deethanizer, and splitter columns, to recover hydrogen- and methane-rich gas streams, purify olefin products, and separate heavier hydrocarbon fractions.

Due to the scale, strong process coupling, and interconnected nature of these operations, olefins plants typically experience a variety of instability events during online operations. These events may be broadly categorized into:

\begin{itemize}

\item {\textbf{Minor events}}: These events refer to process instabilities such as fluctuating operating pressures, temporary variations in coldbox temperature, rate cuts, and rapid changes in ramp rates. Such disturbances occur without prior notice during online operation and may be driven by upstream variability, equipment interactions, or subtle thermal and hydraulic effects.

\item {\textbf{Major events}}: These events are generally represented by planned shutdown, major instability leading to an unplanned shutdown, and startups. Such transitions often involve manual operator action as well as large magnitude changes in several process variables.

\end{itemize}

\begin{figure}[htbp]
    \centering
    \includegraphics[width=1\textwidth]{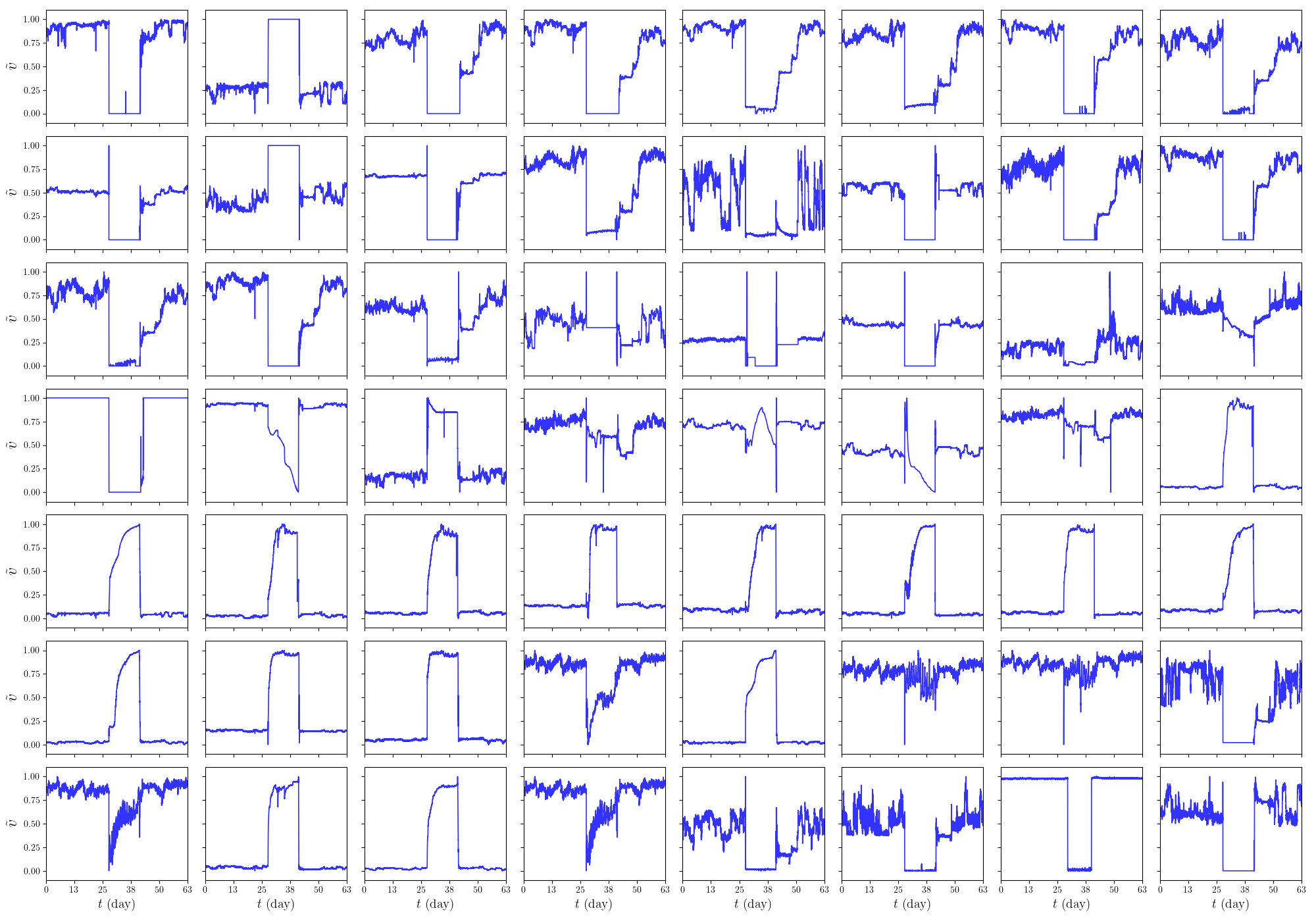}
    \caption{Normalized time-series of the measured process variables $(\tilde{v})$ across the full monitoring horizon of approximately $t=63$ days, with measurements recorded every minute. The multivariate dataset exhibits a wide range of transient behaviors, including fluctuations and sustained operating regime changes across different variables.}
    \label{fig:timeseries_norm_var}
\end{figure}

Figure \ref{fig:timeseries_norm_var} shows the normalized time-series of all measured process variables $(\tilde{v})$ across the full monitoring horizon of approximately $t=63$ days, with measurements recorded every minute. Figure \ref{fig:timeseries_norm_var} highlights the significant dynamic variability of the plant, including abrupt transitions, transient excursions, sustained shifts, and fluctuating operating regimes across different variables, thus representing the challenges in monitoring individual variables separately and making real-time decisions using direct inspection alone.

While major events are conventionally well understood and manually initiated, minor events pose a greater challenge for online monitoring frameworks \cite{poulizos2013faultbook} since they typically evolve over short time scales, are not documented in advance, and often produce weak transient signatures embedded in noisy, high-dimensional sensor data. It is  The detection of such minor instability regions in real time is essential, as even brief deviations can propagate through downstream columns, compressors, or separators, thereby potentially degrading product purity, increasing energy consumption, and triggering unnecessary control interventions \cite{nor2020reviewfault}. It is also not uncommon for minor events to evolve into major events if they are not detected early enough and diagnosed with appropriate measures. Given the complexity of the process and the large volume of multivariate time-series sensor data, this system provides a rigorous testbed for evaluating the ability of different monitoring paradigms to detect both minor and major process instabilities.

Based on plant historian records and expert annotations from our industrial collaborators, {\textit{six}} representative instability regions were identified in the olefins plant dataset. These events correspond to: (i) periods of {\textit{higher coldbox}} temperature, (ii) periods of {\textit{lower coldbox}} temperature, (iii) a prolonged plant {\textit{shutdown}}, (iv) a subsequent {\textit{start-up}} transition, (v) a production {\textit{rate cut}}, and (vi) a {\textit{ramp-up}} in production rates. The corresponding time intervals for each event are summarized in Table \ref{tab:event_times}, along with their durations. For convenience, we subsequently refer to these events as Event 1 - Event 6 throughout the rest of this paper.

Major events such as shutdowns and start-ups are typically planned, operator-driven transitions and thus their timing is relatively well known in practice. In contrast, {\textit{minor}} instabilities, such as hot/cold coldbox behavior, ramp-rate transitions, and rate cuts, occur without prior notice and often manifest as subtle, transient deviations embedded within noisy multivariate measurements during plant operations. The accurate and timely detection of these minor events is therefore critical for enabling proactive decision-making, maintaining product quality, ensuring process safety, and supporting effective control action during online process monitoring. Consequently, these events provide a rigorous benchmark for evaluating the effectiveness of the proposed TDA event-detection framework relative to conventional reconstruction- and trajectory-based process monitoring approaches.

\begin{table}[htbp]
\centering
\caption{Summary of events detected during online operation of the olefins plant.}
\label{tab:event_times}

\begin{tabular}{lllc}
\hline
\textbf{Event ID} & \textbf{Event Description} & \textbf{Timeline (Day / Interval)} & \textbf{Duration (mins)} \\
\hline

Event 1 (E1) & Higher coldbox temperature  & Day 6 - Day 8   & 2{,}064 \\
~       & Higher coldbox temperature  & Day 15 & 800 \\
\\[-6pt]

Event 2 (E2) & Lower coldbox temperature   & Day 21          & 40 \\
\\[-6pt]

Event 3 (E3) & Shutdown                    & Day 28 - Day 42 & 19{,}570 \\
\\[-6pt]

Event 4 (E4) & Start-up                    & Day 42 - Day 53 & 16{,}499 \\
\\[-6pt]

Event 5 (E5) & Rate cut                    & Day 61          & 358 \\
\\[-6pt]

Event 6 (E6) & Ramp rates                  & Day 54          & 235 \\
\hline
\end{tabular}
\end{table}

\section{Results and Discussion} \label{sec:results}

We evaluate the monitoring approaches introduced in Section \ref{sec:method} on the industrial olefins plant described in Section \ref{sec:case_study}. To align with the methodological taxonomy adopted in this work, we first discuss the performance of the {\textit{reconstruction-based}} approaches (PCA and AE), followed by the {\textit{trajectory-based}} approaches. Both reconstruction-based and trajectory-based approaches are evaluated on a moving-window basis. For this case study, each window spans $1$ day with $20\%$ overlap between successive windows. Accordingly, the results are reported in terms of the window-based quantities $\varepsilon_k$, ${\bf z}_k$, or $\chi_k$, while the horizontal axis is mapped back to the corresponding physical time scale of the process (e.g., minutes or days) for interpretability and consistency across all monitoring approaches.

\subsection{Reconstruction-Based Approaches} \label{sec:recon_performance}

For reconstruction-based monitoring, the general workflow involves: i) learning a low-dimensional representation of the system from data obtained during nominal operation, ii) projecting incoming measurements associated with each moving window into the learned latent space, and iii) computing reconstruction errors on a window-basis to identify deviations associated with events. 

\subsubsection{Principal Component Analysis (PCA)} \label{sec:pca_performance}

Figure \ref{fig:PCA_AE_results}a shows the normalized multivariate process trajectories across the full monitoring horizon, with the time intervals corresponding to the different events listed in Table \ref{tab:event_times} highlighted. Figures \ref{fig:PCA_AE_results}b1-b2 present the reconstruction-error profile obtained using PCA.

The PCA monitoring results show that the most prominent excursions in the reconstruction error $(\varepsilon_k)$ occur during the major events, namely, the shutdown (Event 3) and the subsequent start-up transition (Event 4). These events induce sustained and significant deviations in the multivariate process trajectories, which are readily reflected in the PCA latent representation and corresponding Q-statistic. In contrast, the minor events, including higher/lower coldbox temperature excursions, rate cuts, and ramp-rate changes, do not produce clearly distinguishable peaks in the reconstruction error. Overall, PCA is effective in identifying major shifts in the operating regime, but it exhibits limited sensitivity to subtle and minor instabilities embedded in the high-dimensional process measurements.

\subsubsection{Autoencoders (AE)} \label{sec:ae_performance}

\begin{figure}[htbp]
    \centering
    \includegraphics[width=0.98\textwidth]{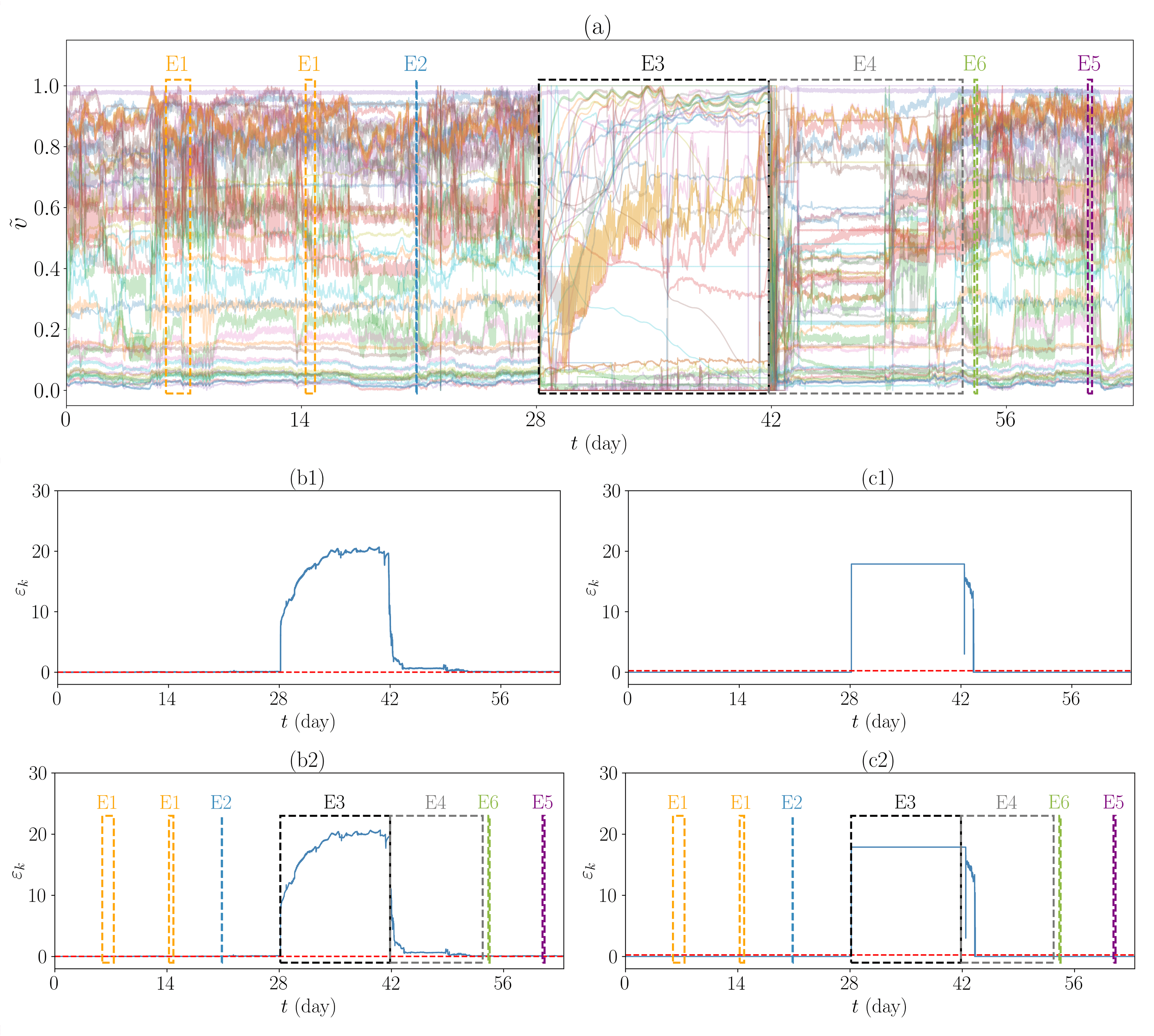}
    \caption{Performance of the reconstruction-based monitoring approaches. (a) Normalized multivariate time-series trajectories $(\tilde{v})$ of the olefins plant over the full monitoring horizon, obtained by superimposing all measured process variables, with the time intervals associated with Events 1-6 highlighted. (b1) PCA reconstruction error as a function of time. (b2) The same PCA reconstruction error overlaid with the event windows described in Table \ref{tab:event_times}. (c1) AE reconstruction error as a function of time. (c2) The same AE reconstruction error overlaid with the event windows described in Table \ref{tab:event_times}. Both reconstruction-based approaches primarily respond to the major events, while exhibiting limited sensitivity to the minor instabilities.}
    \label{fig:PCA_AE_results}
\end{figure}

Figures \ref{fig:PCA_AE_results}c1-c2 present the reconstruction-error profile obtained using the AE model. Similar to PCA, the AE-based monitoring approach responds most strongly to the major events, particularly the shutdown and start-up intervals. Because these events correspond to pronounced process-wide deviations, they are captured by the nonlinear latent representation learned by the encoder-decoder architecture.

However, despite the added representational flexibility of the AE relative to PCA, the reconstruction error $(\varepsilon_k)$ signal remains comparatively insensitive to the minor instabilities. Events 1, 2, 5, and 6 generate weak or no isolated signatures in the AE error profile, making them difficult to detect reliably through reconstruction mismatch alone. Thus, although AE provides a nonlinear alternative to PCA, its detection performance in this case study remains largely concentrated on the major operating transitions. Therefore, the limitations in the ability of reconstruction-based event detection approaches to identify subtle, minor, and dynamically evolving instabilities motivates the use of trajectory-based approaches that explicitly exploit temporal evolution in the latent space rather than relying only on static reconstruction errors.

\subsection{Trajectory-Based Approaches} \label{sec:traj_performance}

For trajectory-based monitoring, the general workflow involves: i) mapping the measurements associated with each moving window into a latent space, followed by ii) analyzing the temporal evolution of the resulting latent variables to identify deviations associated with events. 

\subsubsection{Koopman Autoencoders (KAE)} \label{sec:KAE_performance}

\begin{figure}[htp!]
    \centering
    \includegraphics[width=0.88\textwidth]{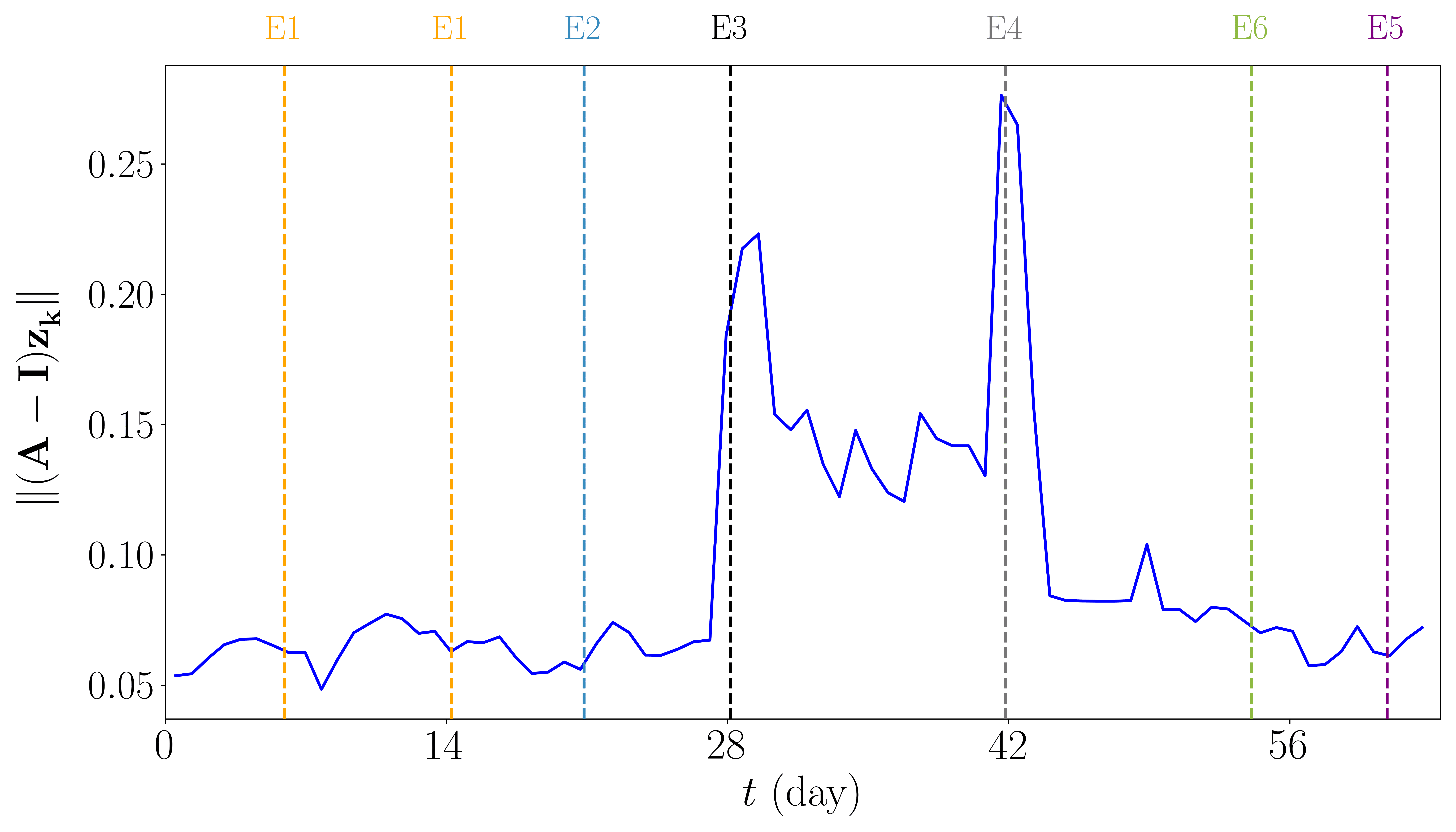}
    \caption{KAE event detection showing the temporal evolution of $\lVert \Delta {\bf z}_k \rVert$ across all time windows overlaid with onset times of the different types of events described in Table \ref{tab:event_times}. Although the latent-space dynamics capture the shutdown and start-up events, the minor instabilities do not produce clearly distinguishable signatures.}
    \label{fig:KAE_results}
\end{figure}

As a representative trajectory-based monitoring approach, KAE infers instability signatures as abrupt changes in process behavior quantified through the discrete latent increment $\Delta {\bf z}_k$ as described in \eqref{error_KAE}. Figure \ref{fig:KAE_results} shows the variation in $\lVert \Delta {\bf z}_k \rVert$ across the sequence of moving windows. The vertical dashed lines in Figure \ref{fig:KAE_results} indicate the onset times of the events described in Table \ref{tab:event_times}. It can be observed that the KAE latent trajectory exhibits its clearest excursions during the major events, particularly in the vicinity of the shutdown (Event 3) and the subsequent start-up transition (Event 4). These operating transitions substantially alter the global plant behavior over extended time intervals, and such changes are reflected in the latent linear dynamics learned by the Koopman framework.

The KAE approach still shows limited sensitivity to the minor events considered in this case study. Events such as higher/lower coldbox temperature excursions, rate cuts, and ramp-rate changes do not generate sufficiently sharp or isolated signatures in $\lVert {\bf ({A}-{I})}{\bf z}_k \rVert$ to enable reliable event detection. Thus, for this industrial olefins data, KAE provides a more dynamical interpretation than reconstruction-based methods, but its practical detection capability still remains largely focused {\textit {only}} on the major instabilities.

An additional practical challenge with KAE models is the selection of an appropriate latent-space dimension. If the latent space is chosen to be too small, important process dynamics may be lost; if it is too large, the learned linear evolution may become more difficult to regularize, interpret, and tune for monitoring purposes. Consequently, these models often require careful calibration to obtain a latent representation that is both dynamically meaningful and effective for event detection. Therefore, the results for this case study indicate that the benchmark approaches for event detection considered in this work still primarily detect the major events and does not consistently identify the onset of the minor instabilities. 

\subsubsection{TDA-NODE Approach} \label{sec:tda_performance}

We next apply the proposed TDA-NODE framework to the same multivariate time-series dataset obtained from the industrial olefins plant. For each moving window, we compute the EC curve associated with the corresponding data matrix using the sublevel-set filtration procedure described in Section \ref{sec:TDA_desc}. This yields a sequence of topological descriptors that compactly summarize the evolving shape of the multivariate process dynamics across time.

An {\textit {important}} empirical observation is that, although TDA does not formally guarantee invariance with respect to variable reordering, the EC and Betti curves obtained in this application exhibit strong robustness to permutations of the row (variable) indices. This behavior can be attributed to the valuation-based, coarse nature of these descriptors, which primarily encode global connectivity statistics of the manifold. Moreover, the topological evolution of the multivariate process appears to be driven predominantly by the time axis rather than by local adjacency in the variable ordering. As a result, reordering sensor variables perturbs the geometry of the manifold but does not induce major changes in the resulting topological summaries.

\subsubsection*{Training Performance}

\begin{figure}[htbp]
    \centering
    \includegraphics[width=0.99\textwidth]{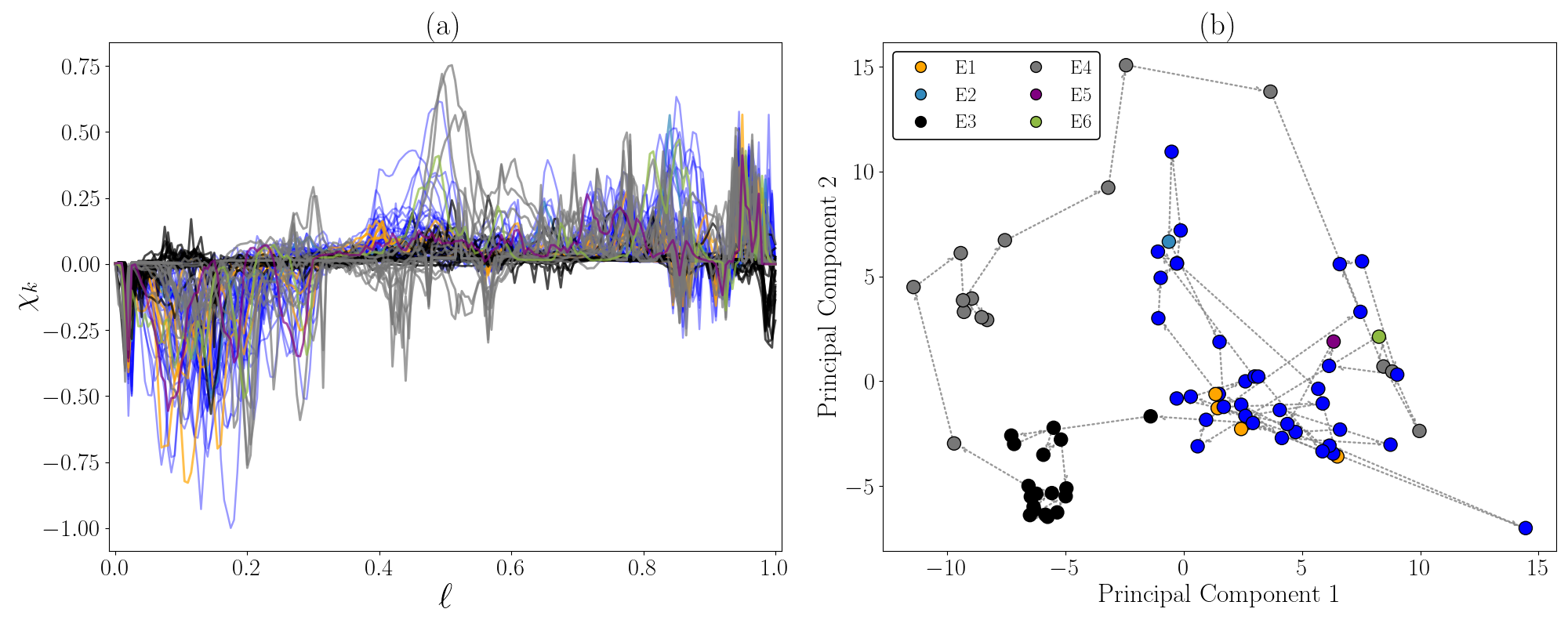}
    \caption{Topological descriptors obtained from the moving-window representation of the olefins plant data. (a) Normalized EC curves computed for all moving windows, color-coded by default operation and event classes as defined in Table \ref{tab:event_times}. (b) PCA projection of the EC curves onto the first two principal components. The minor events mostly overlap with the default operating region, whereas the major events separate more distinctly in the topological latent space. The gray dotted arrowed lines between consecutive PCA markers indicate the dynamic progression of the TDA descriptors $(\chi_k)$, across time windows.}
    \label{fig:EC_PCA}
\end{figure}

Figure \ref{fig:EC_PCA}a shows the normalized EC curves $(\chi_k)$ computed for all moving windows. Each curve captures the evolution of the topology across filtration thresholds $(\ell)$ and thus provides a compact topological summary of the corresponding window data. Figure \ref{fig:EC_PCA}b shows the PCA projection of these EC curves across all moving windows onto the first two principal components, with markers color-coded according to the event labels defined in Table \ref{tab:event_times}. From both the EC curves and their PCA projection, it can be observed that the descriptors associated with the minor events largely overlap with the default operating region. In contrast, the descriptors corresponding to the major events, namely, shutdown (Event 3) and start-up (Event 4), separate more distinctly from the default cluster and trace a clearer excursion in the topological latent space. The gray dotted arrowed lines in Figure \ref{fig:EC_PCA}b indicate the dynamic progression of the EC descriptors across consecutive windows, showing how the process moves away from the default region during shutdown and subsequently transitions back during start-up.

These observations indicate that static clustering of EC curves alone is not sufficient to reliably isolate the minor instabilities from nominal operation. However, the {\textit{directionality}} of the EC descriptors across time contains additional dynamical information that is not captured by the static PCA embedding. To exploit this information, we deploy the NODE model introduced in Section \ref{sec:node} to learn the temporal evolution of the EC curves. In particular, the NODE takes the EC curve for each moving window as the input and predicts its temporal derivative. Similar to the KAE framework, event signatures are therefore inferred from the temporal evolution of the latent representation; however, in the TDA-NODE framework, the topological signatures of events are quantified through the norm $\big\| \dot{\chi}_k \big\|$ across the moving windows.

\begin{figure}[htbp]
    \centering
    \includegraphics[width=0.85\textwidth]{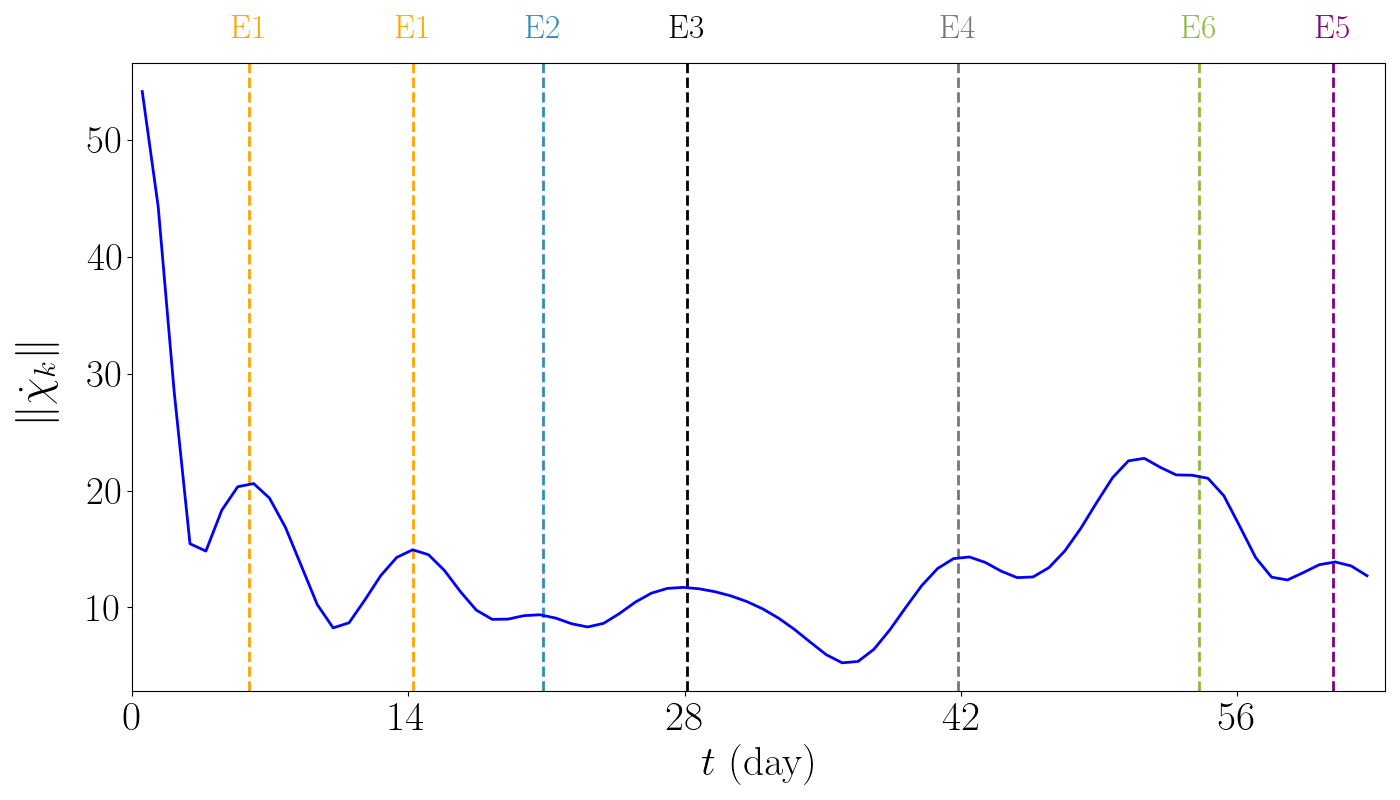}
    \caption{TDA-NODE based event detection showing the temporal evolution of $\big\| \dot{\chi}_k \big\|$ across all moving windows. Colored vertical dashed lines indicate the onset times of Events 1-6 as defined in Table \ref{tab:event_times}. Although the minor events overlap with the default region in the static EC-based latent representation, their signatures are revealed through the directionality of the EC descriptors. Peaks in the predicted topological dynamics align with both minor and major events, indicating that TDA-NODE captures minor and major transitions within a unified trajectory-based monitoring framework.}
    \label{fig:NODE_results}
\end{figure}

Figure \ref{fig:NODE_results} shows the temporal evolution of $\big\| \dot{\chi}_k\big\|$ together with the onset times of the different events defined in Table \ref{tab:event_times}. Unlike the reconstruction-based methods (e.g., PCA and AE), and the trajectory-based KAE model, the proposed TDA-NODE framework exhibits clear excursions not only at the onset of the major events, but also around the minor events. In other words, although the minor events mostly overlap with the default zone in the static EC and PCA views, their effect is reflected in the directionality of the topological descriptors and is captured by the predicted topological dynamics. Thus, sharp variations in $\big\| \dot{\chi}_k \big\|$ provide interpretable and dynamically meaningful signatures for both minor and major instabilities. These results demonstrate that combining TDA representations with a differential ML model enables efficient and sensitive detection of minor as well as major process instabilities directly from high-dimensional multivariate time-series data.

\subsubsection*{Testing Performance}

\begin{figure}[htbp]
    \centering
    \includegraphics[width=1\textwidth]{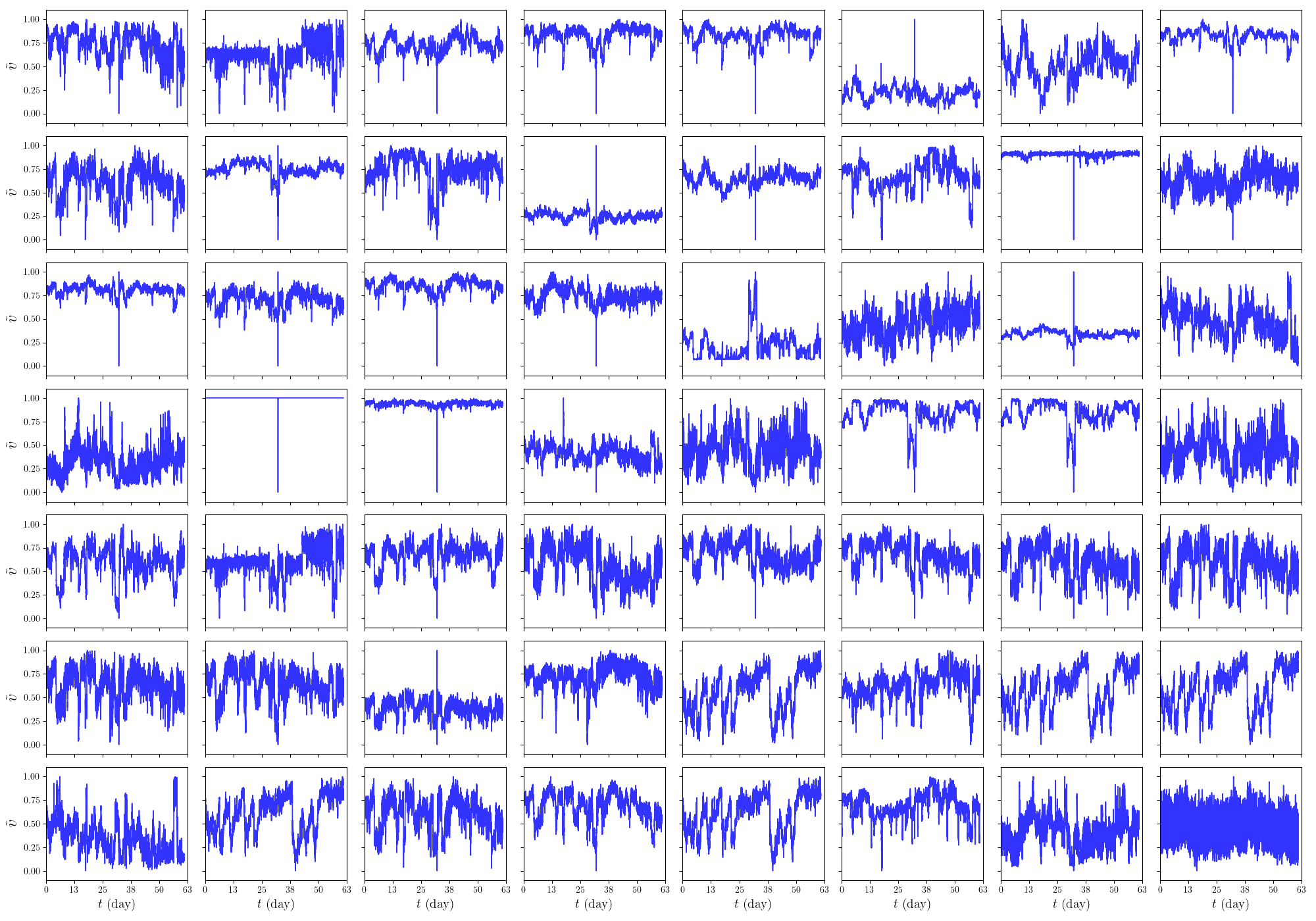}
    \caption{Normalized time-series trajectories of the measured process variables $(\tilde{v})$ for the independent testing dataset. The dataset contains the same process variables as the training dataset, but over a different operating timeline spanning approximately $t=62$ days ($88{,}562$ minutes), with measurements recorded every minute. The trajectories exhibit markedly different transient patterns and dynamic variability relative to the dataset used to develop the NODE model.}
    \label{fig:testing_timeseries}
\end{figure}

To evaluate the generalizability and robustness of the proposed TDA-NODE framework, we deploy the trained NODE model on an independent testing dataset obtained from the same industrial olefins plant. This dataset contains the same process variables as the training dataset, but corresponds to a completely different operating timeline. Measurements are again recorded every minute, spanning approximately $62$ days (i.e., $88{,}562$ minutes). Figure \ref{fig:testing_timeseries} shows the normalized time-series trajectories $(\tilde{v})$ of the testing dataset. Compared with the dataset used to develop the NODE model, these trajectories exhibit markedly different transient patterns, fluctuations, and operating variability, thus providing a meaningful test of whether the learned topological dynamics generalize to unseen plant behavior.

\begin{figure}[htbp]
    \centering
    \includegraphics[width=1\textwidth]{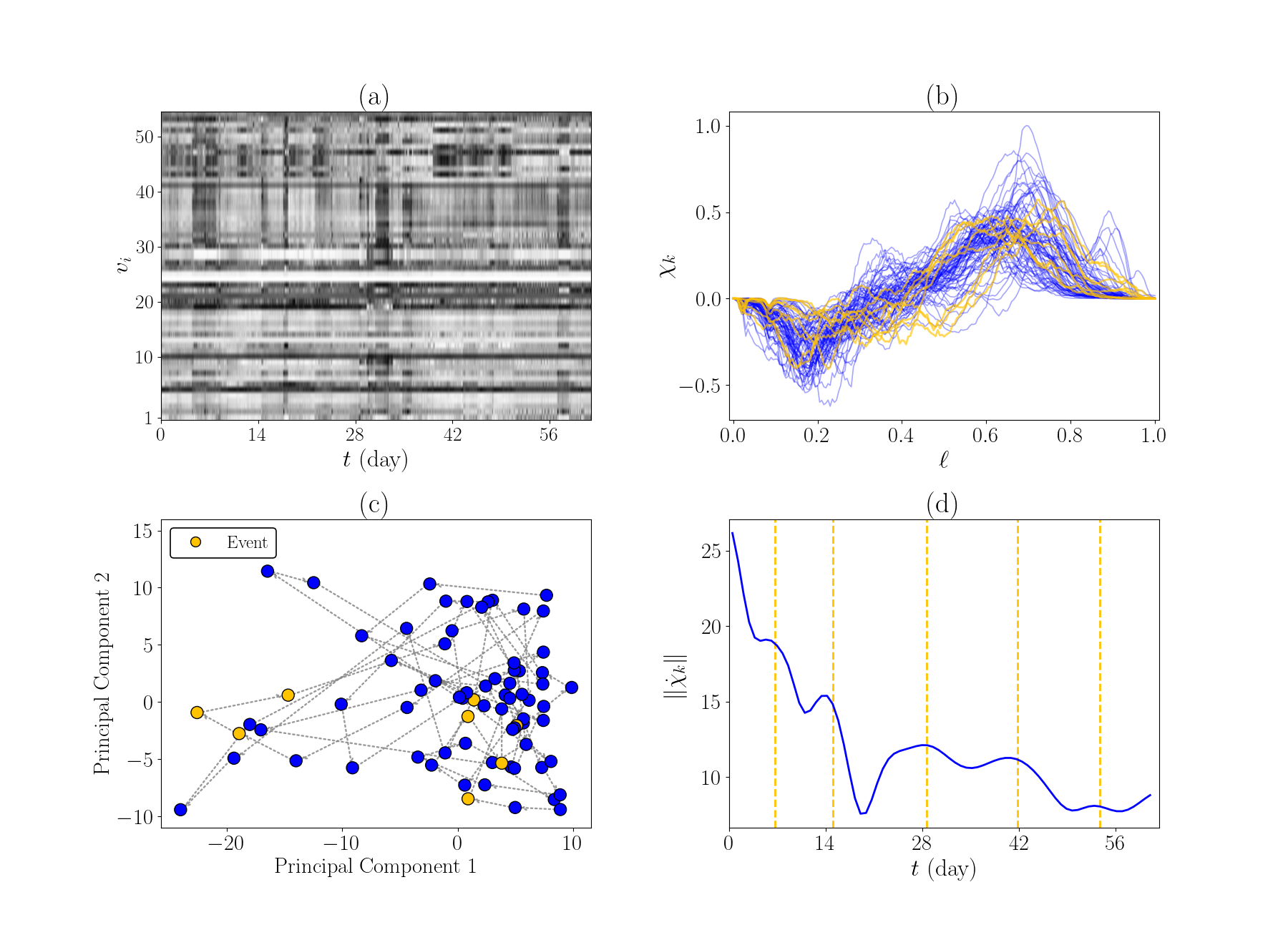}
    \caption{Topological analysis and NODE-based monitoring results for the independent testing dataset. (a) 2D manifold representation of the multivariate time-series data constructed using the same moving-window strategy as in the training case. (b) Normalized EC curves computed for all moving windows of the testing dataset. Although the EC distributions differ from those of the training dataset, the qualitative changes reflect different plant behavior across the two operating periods. (c) PCA projection of the EC curves, where the yellow/orange markers denote events. The events largely overlap with the normal operating region, indicating that static visualization of the topological descriptors alone is insufficient for reliable event detection. (d) NODE-based monitoring of the testing dataset showing the temporal evolution of $\big\| \dot{\chi}_k \big\|$. Peaks in the predicted temporal derivatives align with instability regions in the unseen dataset, demonstrating the generalizability of the pretrained TDA-NODE framework.}
    \label{fig:tda_ec_testing}
\end{figure}

Figure \ref{fig:tda_ec_testing}a presents the 2D manifold representation of the testing dataset constructed using the same moving-window data visualization strategy as in the training case. Figure \ref{fig:tda_ec_testing}b shows the normalized EC curves computed for the corresponding windows. Although the EC distributions for the training and testing datasets are not identical, their qualitative differences reflect variations in actual plant behavior between the two data collection periods. Figure \ref{fig:tda_ec_testing}c shows the PCA projection of the EC curves for the testing dataset, where the events are highlighted. From both the EC curves and their PCA projection, it is evident that the events largely overlap with the normal operating region. Therefore, simply visualizing the low-dimensional topological descriptors is not sufficient to reliably distinguish the occurrence of events in the multivariate time-series data.

To address this limitation, we deploy the optimal NODE model developed using the training dataset. During implementation, the NODE is initialized only with the EC curve corresponding to the first moving window of the testing dataset and is then used in a multi-step-ahead roll-out across the remaining windows without retraining or recalibration. Figure \ref{fig:tda_ec_testing}d shows the temporal evolution of the predicted $\big\| \dot{\chi}_k \big\|$ values for the testing dataset. Although the static EC representations do not clearly separate the events from normal operation, the pretrained NODE still identifies these events through {\textit {distinct peaks}} in the temporal derivatives of the topological descriptors (here, EC curves). This demonstrates that the learned topological dynamics are sufficiently robust to generalize beyond the training period and to detect instability regions in unseen operating data.

These results highlight the practical value of the proposed TDA-NODE framework for online process monitoring. Even when the underlying process trajectories and topological summaries differ substantially from those seen during training, the pretrained model can still identify instability regions through the directionality of the latent topological state. This provides strong evidence that the framework is both robust and generalizable for detecting events in previously unseen multivariate process data.

\section{Conclusions and Future Work} \label{sec:conclusions}

This work presents a process monitoring approach that combines tools from topological data analysis (TDA) and neural differential equations. Starting from high-dimensional multivariate plant data, we represent the process using moving-window manifolds, extract topological descriptors (Euler characteristic curves) through sublevel-set filtration, and learn the temporal evolution of such topological descriptors using neural ordinary differential equation (NODE) models. This yields a compact topological summary of process dynamics that is suitable for online event detection.

The results reveal a clear distinction between the monitoring paradigms. Reconstruction-based methods such as principal component analysis and autoencoders primarily detect the major events, while Koopman autoencoders, although trajectory-based, remains largely sensitive only to the major transitions in this case study. In contrast, the proposed TDA-NODE framework captures both major and minor events by exploiting the directionality of topological descriptors across time. Although the static Euler characteristics of minor events often overlap with the default operating region, their temporal evolution produces clear peaks in the predicted topological dynamics. Moreover, the pretrained NODE model generalizes successfully to an independent testing dataset, demonstrating that the learned topological dynamics are robust to unseen operating conditions.

Future work will focus on extending the framework to additional data representations, such as graph- and correlation-based manifolds, and on coupling topological event signatures with causal analysis and diagnosis tools, along with analysis of additional process systems. A further direction is the integration of topology-based monitoring with control and optimization frameworks to enable real-time data-driven anomaly-detection in industrial process systems.

\section{Acknowledgments} \label{sec:acknowledgement}

We acknowledge the support of the members of the Texas-Wisconsin-California Control Consortium and partial support from the National Science Foundation under grant CBET-2315963.

\bibliography{ref}
\bibliographystyle{unsrt}

\end{document}